%% file: main.tex
\documentclass[twocolumn]{autart}    

\usepackage{graphicx}   
\usepackage{cite}
\usepackage{amsmath,amssymb,amsfonts}

\usepackage{algorithmic}
\usepackage{graphicx}
\usepackage{algorithm,algorithmic}
\usepackage{color}

\input{macros}

\DeclareMathOperator*{\argmin}{arg\,min}

\begin{document}

\begin{frontmatter}

\title{Performance Index Shaping for Closed-loop Optimal Control} 


\author[pur]{Ayush Rai}\ead{rai29@purdue.edu},    
\author[pur]{Shaoshuai Mou}\ead{mous@purdue.edu},               
\author[anu]{Brian D. O. Anderson}\ead{Brian.Anderson@anu.edu.au}  

\address[pur]{School of Aeronautics and Astronautics, Purdue University}  
\address[anu]{School of Engineering, Australian National University} 
          

\begin{abstract}    
The design of the performance index, also referred to as cost or reward shaping, is central to both optimal control and reinforcement learning, as it directly determines the behaviors, trade-offs, and objectives that the resulting control laws seek to achieve.
It is also central to applications such as inverse optimal control and inverse reinforcement learning, where the goal is to infer underlying cost structures and optimal control law from observed behaviors. A commonly used approach for this inference task in recent years is differentiable trajectory optimization, which allows gradients to be computed with respect to cost parameters by differentiating through an optimal control solver. However, this method often requires repeated solving of the underlying optimal control problem at every iteration, making the method computationally expensive. In this paper, assuming known dynamics, we propose a novel framework that analytically links the performance index to the resulting closed-loop optimal control law, thereby transforming a typically bi-level inverse problem into a tractable single-level formulation. Our approach is motivated by the question: given a closed-loop control law that solves an infinite-horizon optimal control problem, how does this law change when the performance index is modified with additional terms? This formulation yields closed-form characterizations for broad classes of systems and performance indices, which not only facilitate interpretation and stability analysis, but also provide insight into the robust stability and input-to-state stable behavior of the resulting nonlinear closed-loop system. 
Moreover, this analytical perspective enables the generalization of our approach to diverse design objectives, yielding a unifying framework for performance index shaping. Given specific design objectives, we propose a systematic methodology to guide the shaping of the performance index and thereby design the resulting optimal control law.
We support our theoretical findings with numerical simulations that demonstrates the utility of our framework in a practical control scenario involving additional design objectives.
\end{abstract}

\end{frontmatter}

\section{Introduction}
\label{sec:intro}
In optimal control theory one deals with constrained optimization of a performance index. In a typical problem, the constraints, or the dynamics, are most often fixed by the physics of the problem. On the other hand, the performance index is most often selected by the design engineer in an attempt to obtain a desired behavior. In practice, the general scheme is to start with a trial performance index and solve the optimal control problem to obtain an optimal control and trajectory. If the response does not match the desired requirements then the entire process is repeated using another performance index, see e.g. \cite[Chapter 2]{lewis2012optimal}. An important concern arises revolving round the requirement of solving the entire optimal control problem again and again. This can well become a very challenging task when dealing with non-linear non-quadratic optimal control problems\footnote{By non-linear non-quadratic optimal control problems we mean optimal control problems where the dynamics are non-linear and/or the performance index is non-quadratic.}. In general, directly designing cost functions poses a significant challenge and is often regarded as an art form in engineering \cite{recht2019tour}. This difficulty is further compounded when, because a closed-loop control law is being sought,  the existence of a solution to a Hamilton-Jacobi-Bellman (HJB) equation is required.
 
Such changes in the performance index are not only encountered because of the need to meet desired requirements but also arise naturally in applications such as safety-critical control, waypoint tracking, and multi-objective optimal control problems involving conflicting goals, such as balancing the trade-off between overshoot and rise time \cite{bass1966optimal, bernstein1993nonquadratic}, or inverse optimal control problems \cite{ab2020inverse}. A key relevant question in such considerations is to establish whether the optimal control problem arising after addition of additional terms to an initially used performance index can be easily solved using some simple adjustment to the solution of the initially solved problem, as opposed to  having to re-solve the adjusted optimal control problem so to speak from scratch. 

\subsection{Related work}

In recent years, there has been increasing interest in end-to-end learning approaches to optimal control, giving rise to the area of differentiable trajectory optimization \cite{bahl2020neural, jin2020pontryagin, amos2018differentiable}. In classical optimal control and robotics, the control policy determines actions at each time step based on the current state, whether through analytical solutions or learned policies. End-to-end learning, by contrast, directly parameterizes and optimizes entire trajectories as continuous functions, shifting the optimization from the control space to the trajectory space itself. Rather than computing control inputs sequentially at discrete time steps, this approach optimizes the full trajectory parameters simultaneously to minimize a trajectory-level objective. This methodology is typically formulated as a bi-level optimization problem, where an inner-loop optimal control problem must be repeatedly solved for each candidate set of trajectory parameters during the outer-loop learning process.

The idea of shaping the cost function in optimal control or the reward function in reinforcement learning has been widely explored in the literature, more extensively in the latter. The key insight is that the reward or cost function used to guide an agent or controller does not need to directly match the true task objective, yet it can still lead to high performance with respect to that objective \cite{singh2009rewards}. In \cite{jain2021optimal}, the authors argue that, because MPC has inherent limitations such as short planning horizons, local optima, and approximate dynamics, optimizing the true cost can sometimes produce suboptimal behavior during execution. Their goal was therefore to redefine which cost functions MPC should optimize so that the resulting rollouts achieve lower true cost despite these limitations. In the context of reinforcement learning, reward shaping techniques have been used for multiple purposes. For example, \cite{sorg2010reward} formulates an optimal reward problem aimed at finding an agent reward function that maximizes the designer’s objective reward, while other work has shown that reward shaping can make the learning process more efficient without altering the optimal policy \cite{gupta2022unpacking}. In our work, we focus on performance index shaping to either achieve or learn desired behaviors while providing analytical tractability. In both cases, these modifications lead to changes in the resulting optimal control law of the corresponding optimal control problem. 

\subsection{Contribution}

In this work, we address the problem of performance index shaping in closed loop infinite horizon optimal control. Starting from a known solution, to an existing optimal control problem, i.e. a specified feedback law,  we introduce additional tunable terms into the performance index and analyze how these modifications influence the resulting optimal control law. By establishing a direct analytical relationship between the performance index and the corresponding closed loop solution (i.e. the control law), our approach collapses what is typically a bi-level optimization problem, where the outer problem adjusts the performance index and the inner problem solves for the optimal control law, into a single level formulation that is both analytically and computationally more tractable. This reformulation eliminates the need to repeatedly solve the optimal control problem from scratch when adapting to new performance objectives. Building on this analytical foundation, we propose an iterative gradient based algorithm for performance index shaping in trajectory optimization, which systematically modifies the performance index while efficiently updating the control law using the derived relationship.

When the original control problem is an LQR problem, the inclusion of additional polynomial terms of higher order than quadratic in the performance index was dealt with in \cite{rai2024nonlinear,bass1966optimal,bernstein1993nonquadratic}. The key application explored in these papers was balancing the tradeoff between the overshoot and the rise time for linear systems. It was demonstrated that the addition of higher-order terms in the LQR performance index can result in controllers with a unique behavior where the control gain increases with larger deviations from zero state and decreases with smaller deviations. These larger control gains reduce the rise/decay times without necessarily affecting the overshoot. The problem developed here deals with the case when the original control problem itself may include both a non-linear system and non-quadratic performance index. 

Our contributions in more specific terms are as follows:
\begin{itemize}
    \item We develop an analytical framework that links changes in the performance index to the resulting closed loop optimal control law, starting from a known solution. The changes are \textit{not} necessarily small, but obey certain restrictions.  This reformulation collapses the typically bilevel structure, where the performance index is tuned through an outer optimization and the control law is computed through an inner optimization, into a single level problem. We show that the  closed-loop nonlinear system resulting from change of the performance index is globally asymptotically stable and input-to-state stable.
    \item We propose a comprehensive framework for performance index shaping. Given specific design objectives, our approach systematically guides the optimal shaping of the performance index and the design of the resulting optimal control law. We demonstrate the applicability of this framework to both linear and nonlinear systems.
\end{itemize}

The article is structured into seven sections. In Section \ref{sec:problem_formulation}, we present the problem formulation. For a given closed-loop optimal control problem, we investigate how to solve the modified optimal control problem with additional terms in the performance index without having to resolve the entire problem from scratch. In Section \ref{sec:analytical}, we propose an analytical approach (building upon \cite{bass1966optimal} and \cite{rai2024nonlinear}) to address this problem, accompanied by relevant theoretical results. In Section \ref{sec:framework}, given a set of design requirements, we develop a performance index shaping algorithm using the framework established in Section \ref{sec:analytical}, which can be directly applied to any solved closed-loop optimal control system. We present various numerical simulations in Section \ref{sec:applications} to demonstrate the performance and applicability of our developed algorithms. Finally, we conclude the paper and outline future directions in Section \ref{sec:conclusion}.
\section{Problem formulation}
\label{sec:problem_formulation}

In this work, we consider a continuous time-invariant non-linear control affine system 
\begin{equation}\label{eq:dynamics}
    \dot x=f(x)+g(x)u, \quad x(0)=x_0,
\end{equation}
where $x \in  \R^n$, $f(x):\R^n \rightarrow \R^n$, $u  \in \R^{p}$, and $g(x):\R^n \rightarrow \R^{n\times p}$. We make the following assumptions about the system described by \eqref{eq:dynamics}:
\begin{assumption}
\label{assumption:dyn}
    (1). The functions $f(x)$ and $g(x)$ are assumed to be smooth, with $f(0)=0$.
    (2). The system is assumed to have the controllability property: for some $T > 0$ and for any initial state $x(0)$, there exists a control input $u(t)$ defined for $0 \leq t \leq T$ such that $x(T) = 0$.
\end{assumption}

We define a \emph{nominal} performance index
\begin{equation} 
\label{eq:nomial_pi} V_0\left(x_0,u(\cdot)\right)=\lim_{T\to\infty}\int_0^T[\|u(t)\|_R^2+m_0(x)]dt,
\end{equation}
where $R \in \R^{m\times m}$ is a positive definite matrix and $m_0(x)$ where $m_0(x)=x^{\top}Qx+q(x)$, $Q$ is non-negative definite symmetric, and $q(x)$ is non-negative, smooth, and of higher order than quadratic in a neighborhood of the origin. This makes $m_0$ a non-negative definite, radially increasing function that is convex (i.e., has a non-negative definite Hessian) at the origin.

The following assumption serves to assure that the optimal system will be locally exponential stable around the origin, and is justified by standard LQ regulator theory \cite[Chapter 3]{anderson1971optimal}.

\begin{assumption}
\label{assumption:obs}  
For the linearized system of \eqref{eq:dynamics} around the origin, we assume that the pair $\left[\frac{\partial f}{\partial x}(0),g(0)\right]$ is controllable in the standard linear system theory sense, and that the pair $\left[\frac{\partial f}{\partial x}(0), Q^{1/2} \right]$ is completely observable.
\end{assumption}

\begin{definition}
    Admissible control law: Control laws are considered admissible if they asymptotically stabilize the system, i.e., $x(t) \to 0$ as $t \to \infty$, and yield a finite integral in \eqref{eq:nomial_pi} for all $x(0)$.
\end{definition}

Using the Hamilton-Jacobi theory, the solution to this optimal control problem, i.e. an admissible optimal control minimizing the performance index \eqref{eq:nomial_pi} subject to system dynamics \eqref{eq:dynamics} is given by
\begin{equation}\label{eq:oclaw_nominal}
    u_0^*=-\frac{1}{2}R^{-1}g(x)^{\top}\nabla \phi_0(x),
\end{equation}
where $\phi_0(x)$ denotes the minimum performance index that under Assumption \ref{assumption:dyn}, satisfies the steady-state Hamilton-Jacobi-Bellman (HJB) equation \cite[Chapter 2]{anderson2007optimal}
\begin{align}
[\nabla\phi_0(x)]^{\top}f(x)&-\frac{1}{4}[\nabla\phi_0 (x)]^{\top}g(x)R^{-1}g(x)^{\top}\nabla\phi_0(x)\nonumber\\&+m_0(x)=0.
\label{eq:HJlinF}
\end{align}

Next, to introduce adaptability in the performance index, we consider a modified optimal control problem with a new performance index that includes an additional function $v(x,\theta)$, which is nonnegative in $x$ for any fixed parameter value $\theta$, added to the nominal loss function in \eqref{eq:nomial_pi}. The new index is given by
\begin{equation} 
\label{eq:new_pi} V_\theta\left(x_0,u(\cdot)\right)=\lim_{T\to\infty}\int_0^T[\|u(t)\|_R^2+m_0(x) + v(x,\theta)]dt,
\end{equation}
where $\theta$ is a free tunable parameter. We assume that $v(0,\theta)=0$, and $v(x,0)=0$. This is a weak assumption and is readily satisfied when $v(x,\theta)$ is a homogeneous multinomial in the entries of $x$ and coefficients $\theta$ or is a neural network with zero bias. Subsequently, we shall indicate restrictions on the functional form of $v$, which still allow significant freedom.

\begin{remark}
    The proposed performance index \eqref{eq:new_pi} introduces adaptability to the system through the parameter $\theta$, which can be freely tuned. Solving \eqref{eq:new_pi} can thus be interpreted as finding the optimal control law for a class of closed-loop optimal control problems.
    
\end{remark}

Next, we consider a single measure of achievement of the performance objectives, which represent the high-level goals we aim to achieve on the resulting optimal trajectory. Here, we distinguish between the performance objectives (or design objectives) and the performance index: the former defines the desired outcomes we want to achieve, while the latter is the cost function of the optimal control problem that is solved to generate the feedback control law. In the first instance, this measure is computed using a given trajectory, or more generally a finite set of trajectories, along with the corresponding feedback control law intended for use on arbitrary trajectories. Each trajectory naturally arises from a specific initial state, and each set of trajectories from a set of initial conditions. Importantly, while these measures evaluate how well the performance objectives are met, the control law itself is expressed as a feedback law, rather than a time-dependent function.

Let $\zeta = \{x(t,\theta),u^*(x(t),\theta) | \;\forall \;t\geq 0\}$ refer to a specific trajectory (states and control inputs) of the dynamics governed by equation \eqref{eq:dynamics}, with control law minimizing \eqref{eq:new_pi}. Let $L(\zeta(\theta))$ represent the design objective encapsulating some given design requirements for that trajectory. This is a (scalar) cost functional tailored to capture various design specifications, such as balancing the trade-off between rise time and overshoot, achieving human-defined performance goals, trajectory tracking, and even enforcing constraints. Minimization through choice of $\theta$ corresponds to some kind of best achieving of these specifications.  When design requirements are expressed as constraints, a barrier function approach can be employed as part of the formulation of the appropriate design objective function. Note that in contrast to the way the parameter $\theta$ arises in the loss function of performance index, being associated just with the state terms in the loss function, the design objective functional can reflect $\theta$-dependence via the input, including as it does the input value given by the law when operating with a particular initial state. Also, a single functional can be used to express multiple design objectives, just as a single performance index in a sense can be used to capture multiple objectives.

Next, suppose that there are $k$ specific trajectories $\zeta_i$ corresponding to $k$ different (but representative) initial states. We capture this wider notion by replacing $L(\zeta(\theta))$ by $\sum_iL(\zeta_i(\theta)$.
By using representative trajectories or equivalently initial states, we seek to tune $\theta$ to minimize this aggregate performance measure to give a closed-loop control applicable to all trajectories and initial states.  

Our overall problem is to design an optimal control law that minimizes or satisfies these design specifications expressed through $\sum_i L(\zeta_i(\theta))$. We shall do this by adjusting the performance index parameters that give rise to the law, with the adjustment process reflecting the effect of adjustment on the value taken by the design objective functional $L$.  In essence, using the parameter $\theta$, we are learning  approximation of $v(x,\theta)$ that is being used as surrogates to effectively model $L$ as the measure of adherence to design objective.

The objective can be simply understood as:
\begin{align}
    &\min_\theta \sum_iL(\zeta_i(\theta)),\\
    \text{where} \quad\quad & \zeta_i = \{x(t,\theta),u^*(x(t),\theta) | \;\forall \;t\geq 0\}, \nonumber\\
     \quad\quad & u^*(x,\theta) \in \argmin_{u(\cdot)} V_\theta(x_0^i,u(\cdot)),\nonumber\\
    \text{subject to} \quad & \dot x=f(x)+g(x)u;\; x(0) = x_0^i.\nonumber
\end{align} 

\begin{remark}
    We emphasize that $L$ cannot be directly treated as the performance index for an optimal control problem, because it is evaluated over a single trajectory (or a finite set of trajectories) rather encompassing all possible initial states, and also that various assumptions on the cost function are needed for the existence of a solution to the Hamilton-Jacobi-Bellman (HJB) equation (continuity, non-negative definiteness,  convexity, etc). 
\end{remark}

\section{Modification of nominal optimal control law using additional terms}
\label{sec:analytical}

Given the nominal minimum performance index $\phi_0$ and the corresponding optimal control law $u_0^*$ for a nominal optimal control problem, our objective is to determine an admissible control law that minimizes a new performance index $V_\theta$, defined in \eqref{eq:new_pi}.  
In addition, we aim to design a structured additional function $v(x, \theta)$, which enables the analytical characterization of the new optimal control law by exploiting the known structure of the nominal solution $(u_0^*, \phi_0)$.

A conventional approach to obtain the new optimal control for a given \( v(x, \theta) \) would involve solving the Hamilton-Jacobi-Bellman (HJB) equation associated with the modified performance index, potentially using the nominal control law \( u_0^* \) as a warm start. However, this process must be repeated each time a new function \( v(x, \theta) \) is proposed, making it computationally inefficient for design and learning tasks that involve iterating over multiple candidates.

To overcome this issue, we propose a specific structure for \( v(x, \theta) \), constructed by additively decomposing it into two terms with the help of functions \( \bar{m}(x, \theta) \) and \( h(x, \theta) \), such that the new performance index admits an analytically tractable value function and its corresponding optimal control law. The function $\bar m(x,\theta)$ has great freedom, while the function $h(x,\theta)$ depends on it, as described below. This structured choice of \( v(x, \theta) \) allows us to directly exploit the known solution \( (u_0^*, \phi_0) \) of the nominal problem and obtain closed-form expressions for the new quantities of interest. We note that this general idea builds on the work of Bass and Webber \cite{bass1966optimal}; however, our approach is more general, as it extends to systems that are affine but potentially nonlinear with non-quadratic costs, rather than being restricted to the linear-quadratic setting considered in their work.

We propose the following structured form for the additional cost function \( v(x, \theta) \):
\begin{align}
v(x, \theta) = \bar{m}(x, \theta) + \frac{1}{4} \left\| R^{-1} g(x)^\top \nabla h(x, \theta) \right\|_R^2,  
\label{eq:proposed_v}
\end{align}
where \( \bar{m}(x, \theta) \) and \( h(x, \theta) \) are non-negative, smooth functions that vanish at $\theta = 0$, and are related as shown below.

The proposed performance index is given by
\begin{align}
V_\theta\left(x_0,u(\cdot)\right)=&\lim_{T\to\infty}\int_0^T \Big[\|u(t)\|_R^2+m_0(x) + \bar m(x,\theta) \nonumber\\ &+\frac{1}{4}\|R^{-1}g(x)^\top\nabla h(x,\theta)\|_R^2\Big]dt,
\label{eq:proposed_pi} 
\end{align} 
where $h(x,\theta)$ is defined by
\begin{equation} 
\label{eq:the_condition} 
\nabla h(x,\theta)^\top
    \left(f(x)-\frac{1}{2} g(x) R^{-1} g(x)^\top \nabla \phi_0(x)\right) = -\bar m(x,\theta),
\end{equation}
where $\phi_0$ is the minimum performance index for the nominal performance index \eqref{eq:oclaw_nominal}.

\begin{theorem}
\label{thm:optimality}
Consider the system \eqref{eq:dynamics} and the nominal performance index in \eqref{eq:nomial_pi}, under Assumptions \ref{assumption:dyn} and \ref{assumption:obs}. With $\phi_0$ as the minimum performance index of the nominal optimal control problem, the minimum performance index of the proposed optimal control problem \eqref{eq:proposed_pi} with non-negative smooth functions $h(x,\theta)$ defined by \eqref{eq:the_condition} is given by 
\begin{equation}
\phi_p(x,\theta)=\phi_0(x)+h(x,\theta),
\label{eq:ana_perf_index}
\end{equation} 
while the optimal control law is given by
\begin{align}
    u_p^*=-&\frac{1}{2}R^{-1}g(x)^{\top}\nabla\phi_p(x,\theta)\nonumber\\ = -&\frac{1}{2}R^{-1}g(x)^{\top}\nabla\phi_0(x)-\frac{1}{2}R^{-1}g(x)^{\top}\nabla h(x,\theta).
    \label{eq:ana_opt_control}
\end{align}
The resulting closed-loop system with the optimal control law is globally asymptotically stable.
\end{theorem}

{\bf{Proof:}} 
We first show that the proposed performance index (with associated control law) satisfies the steady state Hamilton-Jacobi-Bellman (HJB) equation. However, satisfying the HJB equation alone is not sufficient but only necessary to guarantee optimality of the index and the law. Recall that in the linear-quadratic (LQ) case, the associated algebraic Riccati equation may admit multiple solutions, only one of which corresponds to the true optimal performance index and control law, and the quadratic form associated with each solution is a solution of the steady state HJB equation. Likewise then, in the nonlinear case, we must ensure that we have the correct solution of the HJB equation.  To uniquely identify the optimal solution, we will additionally invoke stability properties of the closed-loop system. These stability guarantees ensure that the candidate value function indeed corresponds to the minimum achievable cost and that the resulting control law is admissible. 

Using the definition \eqref{eq:ana_perf_index} and \eqref{eq:the_condition} one can simplify
\begin{align*}
    \nabla&\phi_p(x,\theta)^\top f(x)-\frac{1}{4}\nabla\phi_p (x,\theta)^\top g(x) R^{-1} g(x)^\top \nabla \phi_p(x,\theta) \nonumber\\ 
    =& \nabla(\phi_0(x)+h(x,\theta))^\top f(x) \nonumber\\ &-\frac{1}{4}\nabla(\phi_0(x)\!+\!h(x,\theta))^\top g(x) R^{-1} g(x)^\top \nabla(\phi_0(x)\!+\!h(x,\theta)) \nonumber\\ 
    =& \nabla\phi_0(x)^\top f(x)-\frac{1}{4}\nabla\phi_0 (x)^\top g(x) R^{-1} g(x)^\top \nabla \phi_0(x) \nonumber\\
    & + \nabla h(x,\theta)^\top
    \left(f(x)-\frac{1}{2} g(x) R^{-1} g(x)^\top \nabla \phi_0(x)\right) \nonumber\\
    & -\frac{1}{4}\nabla h(x,\theta)^\top g(x) R^{-1} g(x)^\top \nabla h(x,\theta) \nonumber\\
    = & -m_0(x) -\bar m(x)-\frac{1}{4}\|R^{-1} g(x)^\top \nabla h(x,\theta)\|_R^2 \numberthis
    \label{eq:phi_p_deri}
\end{align*}
The right side is the negative of the state-only terms in the cost function appearing in the proposed performance index \eqref{eq:proposed_pi}; hence the Hamilton-Jacobi equation is satisfied. 

To verify the stability property one studies the closed loop using the optimal control law $u_p^*$:
\begin{align*}
    \dot x  = f(x)-\frac{1}{2}g(x)R^{-1}g(x)^{\top}\nabla\phi_p(x,\theta)
\end{align*}
Since $\phi_p$ (in \eqref{eq:ana_perf_index}) is a radially unbounded positive definite function, we consider it as the Lyapunov function. Taking its derivative and using \eqref{eq:phi_p_deri} results in
\begin{align*}
    \dot \phi_p(x,\theta) =& \nabla\phi_p(x,\theta)^\top \left( f(x)\!-\!\frac{1}{2}g(x)R^{-1}g(x)^{\top}\nabla\phi_p(x,\theta)\right),\\
    =& -m_0(x) -\bar m(x,\theta)-\frac{1}{4}\| R^{-1}g(x)^\top \nabla h(x,\theta)\|_{R}^2 \\ & -\|u_p^*\|_R^2, 
\end{align*}
which is negative semidefinite. Hence using the Lasalle theorem \cite{khalil1992nonlinear} as $\dot \phi_p(x,\theta)=0$ only when $x=0$, one achieves global asymptotic stability. 

Using \textit{any} solution of the Hamilton-Jacobi equation, call it $\bar\phi_p(x,\theta)$, it is not hard to show that the finite time index
\begin{align}
V_\theta\left(x_0,u(\cdot),T\right)=&\int_0^T \Big[\|u(t)\|_R^2+m_0(x) + \bar m(x,\theta) \nonumber\\ &+\frac{1}{4}\|R^{-1}g(x)^\top\nabla h(x,\theta)\|_R^2\Big]dt,
\end{align}
can be written  as
\begin{align}
V_\theta\left(x_0,u(\cdot),T\right)\\\nonumber
=&\int_0^T \|u(t)+\frac{1}{2}R^{-1}g(x)^{\top}\nabla\bar \phi_p(x,\theta\|_R^2dt\\\nonumber
&+\bar\phi_p(x_0,\theta)-\bar\phi_p(x(T),\theta). 
\end{align}

From this formula and the stability property of the control law $u^*_p$ associated with $\phi_p$, it is evident that for any $u(\cdot)$ for which $x(T)\to 0$ as $T\to\infty$
\begin{align*}
\lim_{T\to\infty}V_\theta\big(x_0, u(\cdot), T\big) &\geq \phi_p(x_0, \theta)
\end{align*}
and
\begin{align*}
\lim_{T\to\infty}V_\theta\big(x_0, u^*_p, T\big) &= \phi_p(x_0, \theta).
\end{align*}
 This completes the proof. $\hfill\blacksquare$

\begin{remark}
    We note that if the linearized system around the origin is stabilizable and Assumption \ref{assumption:obs} holds, then local exponential stability of the closed-loop system at the origin is guaranteed. Together with global asymptotic stability, this implies that the closed-loop system is exponentially stable over an arbitrarily large bounded set containing the origin. 
\end{remark}

It is well known that if the optimal control law in the standard LQG problem is $u=kx$, then the insertion of a nonlinear monotonic gain lower bounded by $\frac{1}{2}+\epsilon_1$ and upper bounded by $\epsilon_2^{-1}$ for arbitrarily small positive $\epsilon_1,\epsilon_2$  will not destroy asymptotic stability of the closed-loop \cite{anderson1971optimal}. It is shown in \cite{anderson1969stability} that this property carries over and applies to  $\dot x=f(x)+gu$ (where $g$ is constant) as well as the linear version. The reference \cite{bass1966optimal} also points out that for the subclass of systems of the type considered therein (linear open-loop system, and quadratic loss function to which nonnegative quartic and higher order terms are added), the optimal law can be scaled by any positive constant gain greater than 1/2 without loss of stability. We present a generalized version of these results in the following corollary. 

\begin{corollary}
\label{cor:stability}
    Consider the system \eqref{eq:dynamics} with the same assumptions as Theorem \ref{thm:optimality}. If the optimal control law for the proposed performance index \eqref{eq:proposed_pi} is $u_p^*$ (given by \eqref{eq:ana_opt_control}), then the control law $(\frac{1}{2}+\epsilon)u_p^*$ yields global asymptotic stability for any $\epsilon>0$. 
\end{corollary}

{\bf{Proof:}}
The closed loop using the control law $(\frac{1}{2}+\epsilon)u_p^*$ given us
\begin{align*}
    \dot x  =& f(x)-\frac{1}{4}g(x)R^{-1}g(x)^{\top}\nabla\phi_p(x,\theta)\\&-\frac{1}{2}\epsilon g(x)R^{-1}g(x)^{\top}\nabla\phi_p(x,\theta)
\end{align*}
Using $\phi_p$ (in \eqref{eq:ana_perf_index}) as the positive definite Lyapunov function and taking its derivative one obtains
\begin{align*}
    \dot \phi_p(x,\theta) =& \nabla\phi_p(x,\theta)^\top \Big( f(x)-\frac{1}{4}g(x)R^{-1}g(x)^{\top}\nabla\phi_p(x,\theta)\\&-\frac{1}{2}\epsilon g(x)R^{-1}g(x)^{\top}\nabla\phi_p(x,\theta)\Big),\\
    =& -m_0(x) -\bar m(x)-\frac{1}{4}\|R^{-1} g(x)^\top \nabla h(x,\theta)\|_R^2 \\&-\frac{\epsilon}{2}\|R^{-1} g(x)^\top \nabla \phi_p(x,\theta) \|_R^2, 
\end{align*}
where the last equality utilizes the result from \eqref{eq:phi_p_deri}. Since the Lyapunov function is negative semidefinite, using the Lasalle theorem \cite{khalil1992nonlinear} one achieves global asymptotic stability. 
$\hfill\blacksquare$

For the linear open-loop system, the closed-loop optimal control system is not just asymptotically stable, but input-to-state (ISS) stable. This was demonstrated in \cite{rai2024nonlinear} while for a nonlinear open-loop system, the ISS property for the closed-loop optimal system,  as far as the authors are aware, has not been addressed. There is of course considerable literature \cite{agrachev2008input, sontag1989smooth, sontag1995characterizations} dealing with connections between Lyapunov stability and ISS or BIBS (bounded-input, bounded-state stability), and incremental forms of the latter. While the literature is \textit{not} tied to optimal systems,  we can formulate such a specialised result, but we first review two definitions relevant to input-to-state (ISS) stability.

\begin{definition}
    The system \eqref{eq:dynamics} is input-to-state stable is there exist a $\mathcal{KL}$-function $\beta$ and a $\mathcal{K}$-function $\gamma$ such that, for any input in $u\in L^\infty$ and any $x(0)$, there holds
    \begin{align*}
        \|x(t)\| \leq \beta(\|x(0)\|,t) + \gamma(\|u\|_\infty) \;\; \forall \;t,
    \end{align*}
    where $\|u\|_\infty = \sup (\|u(t)\|,t\geq 0)$.
\end{definition}

\begin{definition} \cite{willems2005mechanisms,sontag2008input}
     Consider a dynamical system together with a function $S: \mathbb{R}^n \times \mathbb{R} \rightarrow \mathbb{R}$, called the storage function, which is positive definite, i.e., $S(0) = 0$ and $S(x) > 0$ for all $x \neq 0$—and proper, i.e., $S(x) \to \infty$ as $\|x\| \to \infty$. In addition, let $w: \mathbb{R}^p \times \mathbb{R}^n \times \mathbb{R} \rightarrow \mathbb{R}$ be a function called the supply rate. The system is said to be dissipative with respect to $w$ if: 1. $S(x,t) \geq 0$ for all $(x,t)$, and 2. the dissipation inequality  
     \begin{align*}
        S(x(t_1),t_1) - S(x(t_0),t_0) \leq \int_{t_0}^{t_1} w(x(t),u,t)\, dt         
     \end{align*}
holds for all $t_1 \geq t_0$.
\end{definition}

\begin{theorem}
    Consider the open-loop system \eqref{eq:dynamics} and with associated performance index \eqref{eq:proposed_pi},  under Assumption \ref{assumption:dyn} and \ref{assumption:obs}. Let $\phi_p(x,\theta)$ be the solution of the associated steady-state Hamilton Jacobi equation, giving rise to the optimal control law $u_p^*$ \eqref{eq:ana_opt_control}. Then the closed-loop system to which is appended an external input $v$, viz. 
        \begin{equation}\label{eq:vinput}
            \dot x=f(x)-\frac{1}{2}g(x)R^{-1}g(x)^{\top}\nabla \phi_p(x,\theta)+g(x)v 
        \end{equation}
            is input-to-state stable (ISS). 
\end{theorem}

{\bf{Proof:}} 
    We proved in Theorem \ref{thm:optimality} that the system \eqref{eq:vinput} is $0$-stable with $\phi_p$ as the candidate Lyapunov function. Consider the value of $\phi_p(x)$ a along trajectories of the system \eqref{eq:vinput}. Evidently
    \begin{align*}
        \frac{d}{dt}\phi_p(x,\theta) =&\nabla \phi_p(x,\theta)^{\top}\dot x \\=&\nabla\phi_p(x,\theta)^{\top}f(x)+\nabla\phi_p(x,\theta)^{\top}g(x)v \\&-\frac{1}{2}\nabla\phi_p(x,\theta)^{\top}g(x)R^{-1}g(x)^{\top}\nabla \phi_p(x,\theta)
    \end{align*}
    Using the steady-state Hamilton Jacobi equation, this can be rewritten as 
    \begin{align*}
        \frac{d}{dt}\phi_p(x,\theta)=&-m(x)-\frac{1}{4}\|R^{-1}\nabla\phi_p(x,\theta)^{\top}g(x)\|_R^2 \\&+(\nabla\phi_p(x,\theta)^{\top}g)v 
        \\
        =&-\!m(x) \! - \! \|\frac{1}{2}R^{-1}\nabla\phi_p(x,\theta)^{\top}g(x)\!-\!v\|_R^2 \! + \! \|v\|_R^2, \\ 
        \leq& -\alpha_1(\|x\|) +\alpha_2(\|v\|),
    \end{align*}
    where $m(x) = m_0(x) + \bar m(x)+\frac{1}{4}\|R^{-1}g(x)^\top\nabla h(x,\theta)\|_R^2$ and $\alpha_1, \alpha_2$ are $\mathcal{K}_\infty$ functions. By applying the dissipation-based characterization of the ISS property \cite{sontag2008input} with the supply rate $w(x(t),u,t) = -\alpha_1(\|x\|) +\alpha_2(\|v\|)$, we conclude that $\phi_p$ serves as an ISS-Lyapunov function, thereby proving that the system \eqref{eq:vinput} is input-to-state stable \cite{sontag1995characterizations}. 
$\hfill\blacksquare$

In the remainder of this section, we address two key issues arising from the preceding analysis.

First, one may reasonably ask: why is it relevant to solve this modified optimal control problem, particularly given that the additional term in the performance index appears to be handcrafted? While the modified cost may initially seem artificial, since its components are not directly motivated by a specific design intention, the structure is intentionally crafted to serve a specific purpose. Namely, it enables the derivation of an analytical solution that leverages the known optimal solution of the nominal problem. This handcrafted structure is not arbitrary; rather, it is engineered so that the resulting modified Hamilton-Jacobi-Bellman (HJB) equation admits a closed-form solution. By introducing a term that has a precise algebraic relationship with the nominal control law, we retain analytical tractability while shaping the performance index in a meaningful way. In this sense, the modified problem serves as a surrogate that balances analytical convenience with practical interpretability.

In more detail, consider the relationship between \( \bar{m}(x,\theta) \) and \( v(x,\theta) \). One might initially think that \( \bar{m}(x,\theta) \) alone could serve as the desired additional cost we wish to impose. However, using only \( \bar{m}(x,\theta) \) would not, in general, lead to an analytically tractable solution. To overcome this, we augment \( \bar{m}(x,\theta) \) with an additional term, yielding the structured form of \( v(x,\theta) \) as defined in \eqref{eq:proposed_v}. This specific augmentation is not arbitrary; rather, it is carefully chosen to preserve analytical tractability while implicitly encoding control effort regularization.

In fact, the additional term in the performance index can be interpreted meaningfully. Specifically,
\begin{align*}
    \frac{1}{4}\left\| R^{-1} g(x)^\top \nabla h(x,\theta) \right\|_R^2 = \left\| u_p^* - u_0^* \right\|_R^2,
\end{align*}
where \( u_p^* \) is the optimal control law resulting from the modified performance index, and \( u_0^* \) is the nominal optimal control. Thus, this term imposes a penalty on the deviation from the nominal control law, effectively regularizing the additional control effort. This can be particularly meaningful in practical scenarios where significant deviation from a known or certified baseline control law is undesirable, either due to safety, robustness, or actuation constraints.

Second, we discuss how to solve \eqref{eq:the_condition}, given a known of $\bar{m}(x,\theta)$. One can adopt any function approximation method to model the function \( h(x,\theta) \). These options include, but are not limited to, polynomial approximation, B-splines, NURBS, radial basis functions (RBF), and neural networks (multilayer perceptron). We will discuss polynomial and neural networks approximation in further detail and use them in the numerical simulations.

Using polynomial approximation, we can express \( h(x,\theta) \) as an infinite sum of linearly independent weighted basis functions \( \{\psi_i(x)\}_{i=1}^{\infty} \), each multiplied by their respective coefficients \( \{c_{i}(\theta)\}_{i=1}^{\infty} \). To make the computation manageable, we truncate the infinite series to a finite number of \( N \) terms, leading to the approximation $h(x,\theta) \approx \sum_{i=1}^N c_i (\theta) \psi_i(x)$, with a desired level of precision. Substituting this approximation into \eqref{eq:the_condition} results in
\vspace{-10pt}

\begin{small}
\begin{align}
    \left(\sum_{i=1}^N c_i(\theta) \nabla \psi_i(x)^\top \!\right)
    \Big(f(x)-\frac{1}{2} g(x) R^{-1}& g(x)^\top  \nabla \phi_0(x)\Big)
    \nonumber\\ &
    \!=\! -\bar m(x,\theta).
    \label{eq:h_approx_poly}
\end{align} 
\end{small}

We can employ the collocation method \cite{russell1972collocation} to achieve an effective approximation of the coefficients \( c_i (\theta) \). This technique involves uniformly sampling \( K \) points within the domain \( \Omega \). By evaluating the approximation \( h(x,\theta) \) at these sampled points, we can generate a system of linear equations incorporating the $c_i(\theta)$ as unknowns and derived from equation \eqref{eq:h_approx_poly}. The goal is to find the least squares solution to this system, which will provide us with an optimal set of coefficients \( c_i \) that best fits our polynomial approximation over the selected sample points. 

Another powerful approximation approach involves the use of neural networks. Specifically, we can approximate \( h(x) \) using a multilayer perceptron as \( h(x,\theta) = \text{NN}(x, \theta) \), where \( \theta \) represents the parameters of the neural network. The network can be trained using uniformly sampled \( K \) points in the domain \( \Omega \), and minimizing the loss function
\begin{align}
\label{eq:h_neural net}
    \mathcal{L}_w =& \nabla NN(x,\theta)^\top
    \left(f(x)-\frac{1}{2} g(x) R^{-1} g(x)^\top \nabla \phi_0(x)\right) \nonumber\\&+\bar m(x,\theta)
\end{align}
where $\nabla NN(x,\theta)$ represents the gradient of the neural network. This algorithm falls within the realm of general physics-informed machine learning.

\section{General framework for performance index shaping}
\label{sec:framework}

So far, our focus has been on efficiently computing optimal control laws when the performance index is augmented with an additional term \( v(x,\theta) \), without the need to re-solve the entire optimal control problem. This approach, grounded in analytical constructions, provides an elegant solution to a class of modified problems. However, when it comes to addressing real-world design requirements, this approach reveals two important limitations.

\begin{enumerate}
    \item As discussed in Section~\ref{sec:analytical}, the additional term must reflect the desired design objectives. Consequently, for many practical requirements, it is not straightforward to construct valid function \( \bar{m}(x,\theta) \) that induce the intended behavior.
    
    \item Even if a suitable additional function, whether $v(x,\theta)$ or $\bar m(x,\theta)$, can be designed to achieve the desired behavior, it may not conform to the specific structure required for analytical tractability, as required in Section~\ref{sec:analytical}. In particular, even when a valid function \( \bar{m}(x,\theta) \) is given to achieve the desired objective, the resulting optimal control law includes an additional term (see Equation~\eqref{eq:the_condition}) that emerges from the derivation and is not explicitly readily determined or estimatable in advance. This term might unintentionally alter the system's behavior, potentially undermining the original design goals.
\end{enumerate}

In this section, we shift our perspective and present a more general framework for performance index shaping that directly incorporates a high-level \textit{design objective}.  Importantly, such objectives are not necessarily embedded within the performance index itself but are instead evaluated at the trajectory level, after the control law has been synthesized. Here we formally pose the task of tuning performance index parameters to minimize a high-level design cost functional, under the assumption that each choice of $\theta$ yields an optimal control law that can be computed analytically. This approach offers a principled mechanism for integrating control synthesis with trajectory-level performance evaluation.

With the background of the theory developed in Section \ref{sec:analytical}, we are interested in minimizing the performance index $V_\theta$ as defined in \eqref{eq:new_pi} where $v(x,\theta)$ is defined by \eqref{eq:proposed_v}, $h(x,\theta)$ satisfies \eqref{eq:the_condition}, and $\phi_0(x)$ is the minimum performance index for the nominal problem given by \eqref{eq:oclaw_nominal}. The functional forms of $\bar m$ and $h$ are assumed fixed in advance, while the parameter value $\theta$ is adjustable.  The parameter vector $\theta$ constitutes the collection of performance index parameters; it may simply comprise coefficients of certain polynomials in the state vector. It may not bear any close evident relationship with the performance objective.

We now propose a gradient-descent based numerical algorithm for adjusting the performance index parameters.  We begin with an initial estimate of $\theta$ as $\theta_0$ and use the update rule
\begin{align}
\label{eq:theta_update}
    \theta_{k+1} \!=\! \theta_{k} \!-\! \gamma\sum_i\frac{\partial L(\zeta_i(\theta)}{\partial \theta} \! =\! \theta_{k} \!-\! \gamma\sum_i\frac{\partial L(\zeta_i(\theta))}{\partial \zeta_i} \frac{\partial \zeta_i(\theta)}{\partial \theta} 
\end{align}
where $\gamma$ is the learning rate, and $\frac{\partial L(\zeta_i(\theta))}{\partial \zeta_i}$ is defined using the concept of functional derivatives \cite{frigyik2008introduction}.

The key issue is to handle the evaluation of the gradient of $L$. If \( L(\zeta_i(\theta)) \) is not a differentiable functional of the trajectory, for such cases one can employ any numerical approximation techniques to approximate the gradient  $\frac{\partial L(\zeta_i(\theta))}{\partial \theta},$ such as finite difference methods or derivative-free optimization techniques \cite{larson2019derivative}. On the other hand, if \( L(\zeta_i(\theta)) \) is a differentiable functional of the trajectory, we can analytically investigate the derivative using the chain rule. We observe that $ \frac{\partial L(\zeta_i(\theta))}{\partial \zeta_i}$ is straightforward to compute since we know how \( L \) depends on the trajectory. We will now discuss how to compute $
\frac{\partial \zeta_i(\theta)}{\partial \theta}, \quad \text{i.e.,} \quad \frac{\partial x}{\partial \theta} \quad \text{and} \quad \frac{\partial u^*}{\partial \theta}.$
For some parameter \( \theta_k \) (considering the \( k \)-th iteration of the algorithm), the optimal control law minimizes the parameter-dependent performance index \eqref{eq:new_pi} is given by
\begin{equation}
    u^*(x,\theta) = -\frac{1}{2}R^{-1}g(x)^{\top}\nabla\phi_0(x)-\frac{1}{2}R^{-1}g(x)^{\top}\nabla h(x,\theta).
\end{equation}
Taking the partial derivative with respect to $\theta$ results in 
\begin{equation}
    \frac{\partial u^*(x,\theta)}{\partial\theta} = -\frac{1}{2}R^{-1}g(x)^{\top}\frac{\partial \nabla h(x,\theta)}{\partial\theta}.
    \label{eq:u_theta}
\end{equation}
For $\frac{\partial x}{\partial \theta}$, if we differentiate the trajectory by $\theta$ and using the chain rule we get 
\begin{equation}
    \frac{\partial \dot x}{\partial\theta} =  \frac{\partial f(x)}{\partial x}\frac{\partial x}{\partial\theta}+\frac{\partial g(x)}{\partial x}\frac{\partial x}{\partial\theta}u^* + g(x)\frac{\partial u^*}{\partial\theta};\; \frac{\partial x}{\partial\theta}(0) = 0.
    \label{eq:x_theta}
\end{equation}
This is a simple ODE for \(\frac{\partial x}{\partial \theta}\) with zero initial conditions, and it can be easily solved forward in time.

\begin{remark}
    If the cost function $L$ contains barrier function-like terms, the initial choice of $\theta$ must ensure that $L$ remains finite.
\end{remark}

\begin{remark}
We can also consider scenarios where the initial condition is uncertain and instead described by a distribution \( \rho_0 \). 
In such cases, one can define the aggregate design objective as the expected loss over trajectories initialized from \( \rho_0 \):
\[
    \mathbb{E}_{x_0 \sim \rho_0} \big[ L(\zeta(\theta; x_0)) \big],
\]
where \( \zeta(\theta; x_0) \) denotes the trajectory generated under the optimal control law starting from \( x_0 \). 
In practice, this expectation can be approximated using a finite set of \( k \) sampled initial conditions, leading to the empirical loss
\[
    \frac{1}{k} \sum_{i=1}^k L(\zeta_i(\theta)), \quad x_i(0) \sim \rho_0,
\]
where each trajectory \( \zeta_i(\theta) \) corresponds to an initial state drawn from \( \rho_0 \).
\end{remark}

\section{Numerical Simulations}
\label{sec:applications}
To demonstrate the effectiveness of the proposed Performance Index Shaping (PIS) framework, we study two examples: a simple linear system and the cartpole. The motivation for considering the former is that, even for a simple nominal LQR system, there is no systematic approach to incorporating nonlinear terms into the cost function to achieve a higher objective. In this case, the nominal optimal control law is obtained by solving the Riccati equation. On the other hand, the cartpole has nonlinear dynamics, for which a nominal optimal controller is obtained by training a neural network to represent the minimum performance index.

\begin{example}
First we consider a third-order linear time-invariant (LTI) system characterized by the following state-space dynamics:
\[
\begin{bmatrix}
\dot x_1 \\
\dot x_2 \\
\dot x_3
\end{bmatrix}
= \begin{bmatrix}
0 & 1 & 0 \\
-\omega_n^2 & -2\zeta \omega_n & 1 \\
0 & 1 & -1
\end{bmatrix}
\begin{bmatrix}
x_1 \\
x_2 \\
x_3
\end{bmatrix}
+ \begin{bmatrix}
0 \\
1 \\
0
\end{bmatrix} u,
\]
where \( \omega_n = 2.0 \) is the natural frequency and \( \zeta = 0.1 \) is a low damping ratio. The upper-left \( 2 \times 2 \) block corresponds to a classical second-order underdamped mass-spring-damper system, which inherently exhibits oscillatory behavior and significant overshoot. The third state variable allows us to test the controller's behavior under additional system complexity.
\end{example}

\begin{figure}[t!] 
    \centering
    \includegraphics[width=1\linewidth]{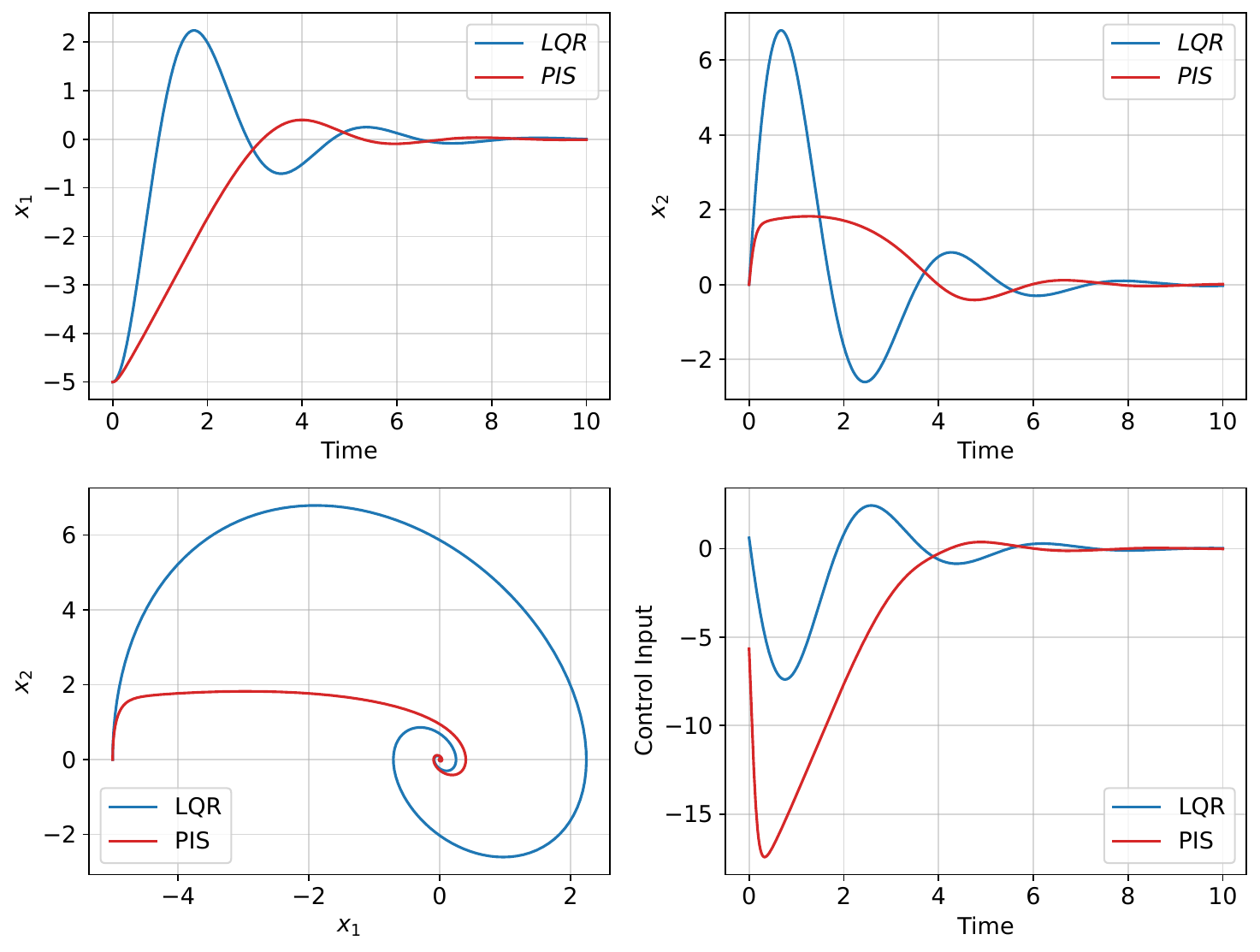}
    \caption{Comparison of closed-loop trajectories under LQR and PIS (Performance index shaping) control for the third-order system. PIS yields improved transient performance by reducing overshoot and accelerating convergence, at the cost of higher control effort.}
    \label{fig:eg1}
\end{figure}

The nominal control objective is formulated through the standard infinite-horizon quadratic cost:
\[
V_0\left(x_0, u(\cdot)\right) = \lim_{T \to \infty} \int_0^T \left[\|u(t)\|_R^2 + \|x(t)\|_Q^2\right] dt,
\]
where \( R = 1 \) and \( Q = I \) (identity matrix), leading to a classic LQR solution. The closed-loop performance of the resulting LQR controller is illustrated in Fig.~\ref{fig:eg1}, showing an underdamped response with noticeable overshoot in \( x_2 \), consistent with the low damping setting.

\begin{figure}[t!] 
    \centering
    \includegraphics[width=0.8\linewidth]{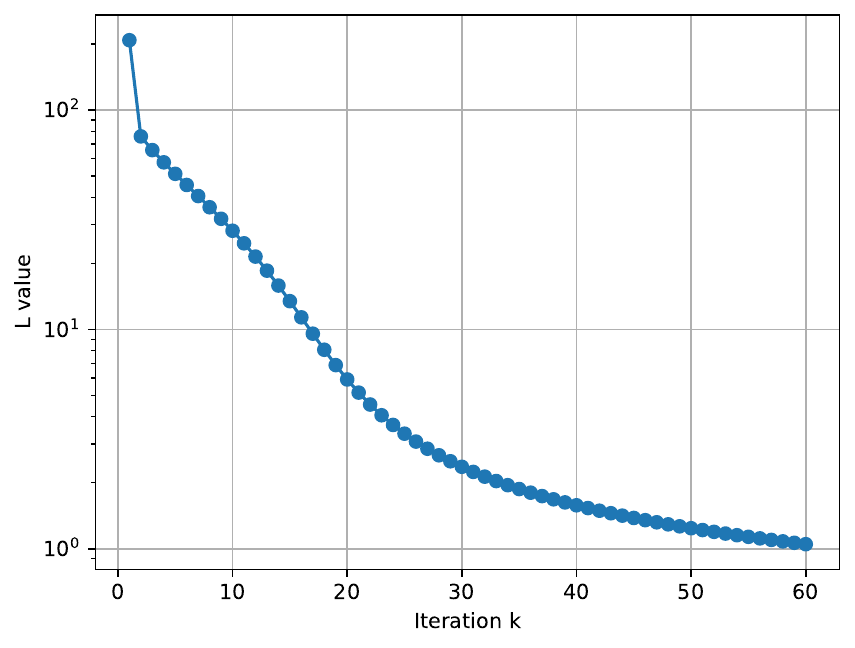}
    \caption{Evolution of the design objective $L$ over iterations for Example 5.1}
    \label{fig:eg1_evolution}
\end{figure}

To improve upon this nominal design, we introduce a high-level design objective that explicitly penalizes overshoot in the second state component, \( x_2 \). Specifically, we define a trajectory-level penalty that becomes active whenever \( x_2(t) \) exceeds a predefined threshold \( x_{\max} \). Let \( \zeta = \{x(t)\}_{t=0}^T \) denote the trajectory resulting from a control law \( u^*(x; \theta) \). The overshoot penalty is expressed as:
\begin{align}
\label{eq:overshoot_1}
    L(\zeta) = \int_{t=0}^T \frac{1}{\beta} \Big( \log\left(1 + \exp\left( \beta \left(x_2(t) - x_{\text{max}} \right) \right) \right)- \log 2 \Big),    
\end{align}
where \(x_{\text{max}} = 2.0\), \( \beta > 0 \) controls the smoothness of the penalty (with larger \( \beta \) approximating a hard ReLU), and \( \log(1 + \exp(\cdot)) \) is the softplus function, a smooth approximation to ReLU. For the purpose of numerical computation, the integral is approximated by a discrete-time summation.
This cost is differentiable with respect to both trajectory and control parameters, making it compatible with learning-based optimization and trajectory shaping. To integrate this design objective into the control synthesis, we augment the original cost with additional terms inspired by \eqref{eq:proposed_pi}, where \( \bar{m}(x, \theta) \) is a tunable shaping term, and \( h(x, \theta) \) satisfies the condition in Equation~\eqref{eq:the_condition}. Both \( \bar{m} \) and \( h \) are parameterized using even polynomial basis functions of degree up to 4 (results in 21 terms), allowing sufficient expressiveness while retaining analytical tractability.

To implement the proposed framework, the parameter vector $\theta \in \mathbb{R}^{21}$ is tuned using a gradient-based update rule \eqref{eq:theta_update}, where the overshoot penalty \eqref{eq:overshoot_1} serves as the design objective and the gradient of the trajectory cost with respect to $\theta$ is computed via sensitivity propagation \eqref{eq:u_theta} and \eqref{eq:x_theta}. Learning is performed on a single trajectory starting from the initial condition $[-5,0,0]$. The Figure \ref{fig:eg1_evolution} shows the evolution of the design objective $L$ over iterations.

Figure~\ref{fig:eg1} shows a comparison between the trajectories resulting from the nominal LQR controller and the PIS-optimized controller. Under LQR, the system stabilizes the states, but exhibits oscillatory transients and overshoot in \( x_2 \), as reflected in the phase portrait and time-series plots. The control input remains moderate, balancing energy and stability. In contrast, the PIS controller yields a sharper transient response with reduced overshoot in \( x_2 \), achieved at the cost of a more aggressive control input. The phase portrait illustrates a faster and more direct convergence to the origin, emphasizing the improved transient performance. This result highlights how PIS allows incorporation of trajectory-level objectives into the control design without sacrificing stability.

\begin{example}
We next consider the nonlinear cartpole system, whose dynamics are described by
\[
\dot{x}(t) =
\begin{bmatrix}
v \\
\omega \\
\dfrac{u + m_p \sin\xi \big( l \omega^2 - g \cos\xi \big)}{m_c + m_p \sin^2\xi} \\
\dfrac{u \cos\xi + m_p l \omega^2 \cos\xi \sin\xi - (m_c + m_p) g \sin\theta}{l \left(m_c + m_p \sin^2\xi \right)}
\end{bmatrix},
\]
where \(x = [p, \xi, v, \omega]^\top\) denotes the cart position, pendulum angle, cart velocity, and angular velocity respectively. The control input \(u\) corresponds to the horizontal force applied to the cart, while \(m_c\) and \(m_p\) are the cart and pendulum masses, and \(l\) is the pendulum length.
\end{example}

The nominal controller is obtained by training a neural network policy to minimize the Hamilton–Jacobi–Bellman (HJB) residual, using the cost
\[
V_0(x_0, u(\cdot)) = \int_0^\infty \left( \|x(t)\|_Q^2 + \|u(t)\|_R^2 \right) dt,
\]
with \(Q = I\) and \(R = 1\). 
This design successfully stabilizes the cart-pole upright equilibrium for initial conditions near the upright position (e.g., $\xi(0) \in [-0.1, 0.1]$ radians). However, under this nominal policy, the pendulum angle $\theta(t)$ can still reach values up to 0.23 radians during transients, which may exceed safe operating ranges. To address this, we design the objective such that the angle remains below $\xi_{\max} = 0.15$ radians at all times.

\begin{figure}[t!] 
    \centering
    \includegraphics[width=0.8\linewidth,trim={0cm 0cm 0cm 0.0cm},clip]{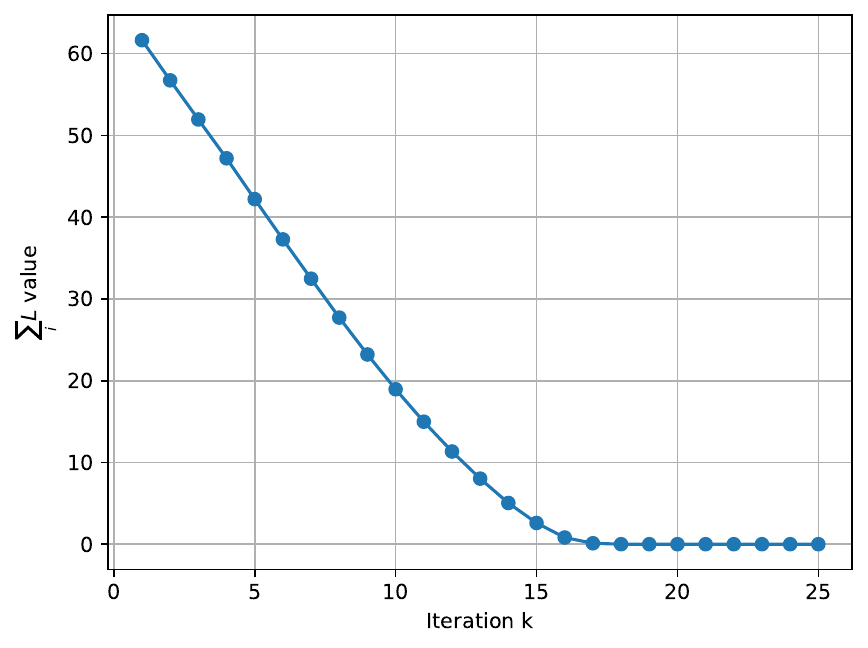}
    \caption{Evolution of the design objective $L$ over iterations for Example 5.2}
    \label{fig:eg2_evolution}
\end{figure}

\begin{figure}[t!]
    \centering
    \begin{subfigure}[b]{0.46\textwidth}
        \centering        \includegraphics[width=0.7\linewidth,trim={0cm 0.3cm 0cm 0.65cm},clip]{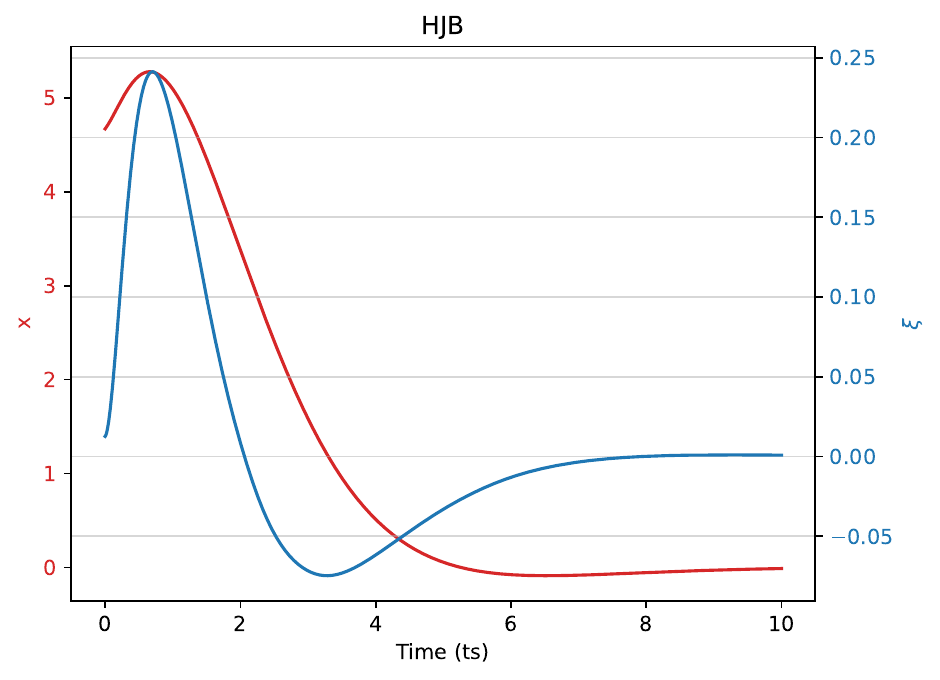} 
        \caption{Nominal trajectory} 
        \label{fig:hjb_cartpole}
    \end{subfigure}
    \hfill
    \begin{subfigure}[b]{0.46\textwidth}
        \centering
        \includegraphics[width=0.7\linewidth,trim={0cm 0.3cm 0cm 0.6cm},clip]{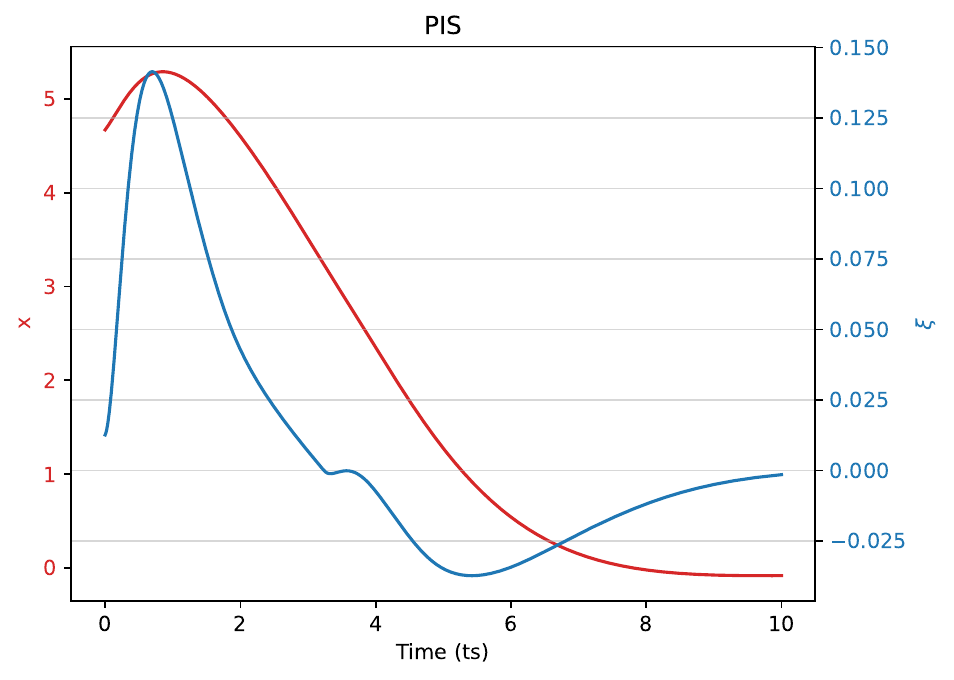}
        \caption{Trajectory tuned using performance index shaping}
        \label{fig:pis_cartpole}
    \end{subfigure}
    \caption{Closed-loop trajectories of the cart-pole under the nominal HJB-based controller and the PIS-optimized (Performance index shaping) controller. The nominal controller stabilizes the upright position but allows pendulum excursions up to \(0.23\) radians. The PIS controller restricts the angle within \(0.15\) radians, improving safety at the cost of higher control effort.}
    \label{fig:cartpole}
\end{figure}

To this end, we introduce a trajectory-level shaping penalty that enforces a soft constraint on the pendulum angle. Specifically, we penalize violations of the threshold \(\xi_{\max} = 0.15\) radians via
\begin{equation}
\label{eq:overshoot_2}
    L(\zeta) = \int_{t=0}^T \frac{1}{\beta} \Big( \log\big(1 + \exp(\beta (\xi(t) - \xi_{\max}))\big) - \log 2 \Big),
\end{equation}
where \(\beta > 0\) regulates the smoothness of the penalty similar to the previous example. For the purpose of numerical computation, the integral is approximated by a discrete-time summation.

In this case as well, the functions \( \bar{m} \) and \( h \) are represented using even polynomial basis functions of degree up to 4, yielding 45 terms in total. The corresponding parameter vector \(\theta \in \mathbb{R}^{45}\) is updated through the gradient-based rule \eqref{eq:theta_update}, with the overshoot penalty \eqref{eq:overshoot_2} defining the design objective. Training is conducted on ten trajectories with initial points uniformly sampled between $[-5, -0.1, -1, -0.1]$ and $[5, 0.1, 1, 0.1]$. The initial condition used to test the new optimal controller is \([4.6, 0.04, 0.9, 0.1]\), the same as in the nominal case. The evolution of the design objective \(L\) over iterations is shown in Figure~\ref{fig:eg2_evolution}.

Figure~\ref{fig:cartpole} compares the closed-loop trajectories. While the nominal controller ensures stabilization, it fails to satisfy the angle constraint. In contrast, the PIS controller yields a sharper transient with reduced angular excursion, highlighting the effectiveness of incorporating trajectory-level design objectives into nonlinear control synthesis.

\section{Conclusions}
\label{sec:conclusion}
This work introduces a novel framework for performance index shaping (PIS) in optimal control, enabling explicit incorporation of high-level design objectives into the cost functional without re-solving the full optimal control problem from scratch. Building on the analytical structure of the Hamilton-Jacobi-Bellman (HJB) equation, we propose a closed-form modification of the performance index that preserves analytical tractability while allowing for flexible trajectory-level tuning. We further demonstrate that the framework accommodates non-standard design criteria, such as overshoot penalties, through parameterized additional functions.

This approach provides a principled and computationally efficient method for co-designing performance indices and optimal control laws. Future directions include extending the framework to constrained systems, model predictive control, and data-driven settings where the nominal controller is learned or approximated from demonstrations.

\bibliographystyle{ieeetr} 
\bibliography{sample}


\end{document}

%% file: macros.tex
\usepackage{caption}
\usepackage{subcaption}

\newtheorem{theorem}{Theorem}[section]

\newtheorem{corollary}[theorem]{Corollary}
\newtheorem{definition}[theorem]{Definition}
\newtheorem{remark}[theorem]{Remark}
\newtheorem{assumption}[theorem]{Assumption}
\newtheorem{example}[theorem]{Example}

\newcommand{\R}{\mathbb{R}}

\newcommand\numberthis{\addtocounter{equation}{1}\tag{\theequation}}